\documentclass[twocolumn,showpacs,superscriptaddress,preprintnumbers]{revtex4}
\usepackage{graphicx}
\usepackage{dcolumn}
\usepackage{bm}

\input{tcilatex}
\begin{document}

\preprint{}
\title{Crystallized merons and inverted merons in the condensation of spin-1
Bose gases with spin-orbit coupling}
\date{Nov 26, 2011}
\author{S.-W. Su}
\affiliation{Department of Physics, National Tsing Hua University, Hsinchu 30013, Taiwan}
\author{I.-K. Liu}
\affiliation{Department of Physics, National Changhua University of Education, Changhua
50058, Taiwan}
\author{Y.-C. Tsai}
\affiliation{Department of Photonics, Feng Chia University, Taichung 40724, Taiwan}
\author{W. M. Liu}
\affiliation{Instuite of Physics, Chinese Science Academy, Beijing 100190, P. R. China}
\author{S.-C. Gou}
\affiliation{Department of Physics, National Changhua University of Education, Changhua
50058, Taiwan}

\begin{abstract}
The non-equilibrium dynamics of a\ rapidly quenched spin-1 Bose gas with
spin-orbit coupling is studied. By solving the stochastic projected
Gross-Pitaevskii equation, we show that crystallization of merons can occur
in a spinor condensate of $^{87}$Rb. Analytic form and stability of the
crystal structure are given. Likewise, inverted merons can be created in a
spin-polarized spinor condensate of $^{23}$Na. Our studies provide a chance
to explore the fundamental properties of meron-like matter.
\end{abstract}

\maketitle

\address{$^{1}$Department of Appled Mathematics, Feng Chia University, Taichung 40724,\\
Taiwan\\
$^{2}$Department of Physics, National Changhua University of Education,\\
Changhua 50058, Taiwan\\
$^{3}$Department of Mathematics, National Taiwan University, Taipei 106,\\
Taiwan}


Spin-orbit coupling (SOC) \cite{Sakurai}\ is an ubiquitous quantum
phenomenon, which links the internal (spin) and orbital (linear or angular
momentum) degrees of freedom of a particle. The best known example of SOC
arises in the motion of electrons in the atom, where the electron's orbiting
around the nucleus can affect the orientation of electron's spin. Recently,
SOC with tunable strength has been realized by Lin \textit{et} \textit{al}. 
\cite{Lin} by shinning two orthogonal laser beams intersecting in a pseudo
spin-1/2 Bose-Einstein condensate (BEC) of $^{87}$Rb \cite{Liu}. The lasers
are detuned from Raman resonance so that the momentum and spin can couple
via exchanging photons in the two orthogonal beams. This opens new
possibilities to simulate the role of SOC for a wide range of phenomena in
condensed matter physics by using ultracold atoms, promising applications to
quantum computing \cite{Bandy}, spintronics devices \cite{Das,Sinova} and
topological insulators \cite{Zhang,Bernevig,Kane}. Inspired by the
experimental realization in Ref. \cite{Lin}, theoretical extensions using
atomic BECs have been studied by a number of authors, including the direct
incorporation of SOC into the spin-1/2, 1 and 2 BECs \cite%
{Hui,Wang,You,Kawakami}, and non-trivial ground state structures have been
predicted. Novel excitations have also been predicted in fermionic gases 
\cite{Yi}, such as the Rashba pairing bound state (Rashbon) \cite%
{Jiang,Jayantha}.

So far, all studies on the ultracold atoms with SOC were focused in the
cases of zero temperature. It is fundamentally important to see how the
nonlocal nature of SOC affects the pattern of spontaneous symmetry breaking.
In this work, we study the non-equilibrium dynamics during the condensation
of a spin-1 Bose gas with SOC. In particular, we focus on the formation of
topological defects in the limit of rapid temperature quench. According to
Kibble-Zurek mechanism \cite{Kibble,Zurek}, topological defects can be
created via phase transitions at finite temperatures, which are caused by
spontaneous symmetry breaking and thermal fluctuations near the critical
point. By solving the stochastic projected Gross-Pitaevskii equation
(SPGPE), we show that, in a spin-1 BEC, the combination of SOC,
spin-exchange interaction and thermal fluctuations\ can generate meron-like
excitations. A meron is a peculiar topological defect that was originally
hypothesized as a half-instanton in the particle physics \cite{Actor} and
later a half-skyrmion in the quantum Hall systems \cite{Brey,Brown}. Since
merons carry half unit of topological charge, it is believed that isolated
merons can only be observed when particular boundary conditions are imposed.
By far, merons have been created in the superfluid $^{3}$He-A \cite%
{Ruutu,Ishiguro} in a rotating cylinder, and the spinor BEC with a special
constraining magnetic field \cite{Ketterle}\textbf{.} In what follows, we
show that, stable collective excitations such as the crystalline orders of
merons and other isolated variants can be created in a rapidly quenched
spinor BEC with SOC.

The order parameter of a spin-1 BEC is given by $\mathbf{\Psi =}\left( 
\begin{array}{ccc}
\Psi _{1}, & \Psi _{0}, & \Psi _{-1}%
\end{array}%
\right) ^{T}$, where $T$ stands for the transpose and $\Psi _{m_{F}}$ $%
\left( m_{F}=\pm 1,0\right) $ denotes the macroscopic wave function of the
atoms condensed in the spin state $\left\vert 1,m_{F}\right\rangle $. The
total particle number, $N$, and total magnetization, $M$, are normalized by $%
\int \left\vert \mathbf{\Psi }\right\vert ^{2}d^{3}r=N$, and $\int \left(
\left\vert \Psi _{1}\right\vert ^{2}-\left\vert \Psi _{-1}\right\vert
^{2}\right) d^{3}r=M$. In the following, we consider SOC of the form $%
H_{so}=\sum_{\alpha }V_{\alpha }\hat{p}_{\alpha }\hat{F}_{\alpha }$, where $%
V_{\alpha }$, $\hat{p}_{\alpha }$ and $\hat{F}_{\alpha }$ are respectively
the coupling strength, the linear momentum and the $3\times 3$ matrix of the
spin-1 angular momentum in the $\alpha \left( =x,y,z\right) $ direction. In
the absence of magnetic field, the dynamics of $\mathbf{\Psi }$ is described
by the following coupled nonlinear Schr\"{o}dinger equations, 
\begin{eqnarray}
&&i\hbar \frac{\partial }{\partial t}\Psi _{j}=\hat{H}_{j}^{GP}\Psi _{j}
\label{GPE} \\
&=&\mathcal{\hat{H}}\Psi _{j}+g_{s}\sum_{\alpha }\dsum\limits_{n,k,l}\left( 
\hat{F}_{\alpha }\right) _{jn}\Psi _{n}\Psi _{k}^{\ast }\left( \hat{F}%
_{\alpha }\right) _{kl}\Psi _{l}  \nonumber \\
&&+\frac{\hbar }{i}\sum_{\alpha }\dsum\limits_{n}V_{\alpha }\left( \hat{F}%
_{\alpha }\right) _{jn}\partial _{\alpha }\Psi _{n}\quad \left(
j,k,l,n=-1,0,1\right)  \nonumber
\end{eqnarray}%
Here $\mathcal{\hat{H}}=-\hbar ^{2}\bigtriangledown ^{2}/2m+U(\mathbf{r})+$ $%
g_{n}\left\vert \mathbf{\Psi }\right\vert ^{2}$ denotes the spin-independent
part of the Hamiltonian, with $U(\mathbf{r})$ being the trapping potential.
The coupling constants $g_{n}$ and $g_{s}$ characterizing the
density-density and spin-exchange interactions, respectively, are related to
the $s$-wave scattering lengths $a_{0}$ and $a_{2}$ in the total spin
channels $F_{total}=0$, $2$ by $g_{n}=4\pi \hbar ^{2}\left(
a_{0}+2a_{2}\right) /3m$, $g_{s}=4\pi \hbar ^{2}\left( a_{2}-a_{0}\right)
/3m $ \cite{Ho,Ohmi}. Note that $g_{n}>0$, whereas $g_{s}$ can be either
positive or negative. A spin-1 BEC is said to be ferromagnetic (FM) when $%
g_{s}<0$, and antiferromagnetic (AFM) when $g_{s}>0$. As we shall focus on
the dynamics of spin texture, we introduce the basis set $\Psi _{\alpha }$ $%
(\alpha =x,y,z)$, such that $\Psi _{\pm 1}=\left( \pm \Psi _{x}+i\Psi
_{y}\right) /\sqrt{2}$ and $\Psi _{0}=\Psi _{z}$. As a result, $\hat{F}%
_{\alpha }\left\vert \alpha \right\rangle =0$, and the spin texture, which
is parallel to the local magnetic moment, is defined by $\mathbf{S}\left( 
\mathbf{r}\right) =i\mathbf{\tilde{\Psi}}^{\dagger }\times \mathbf{\tilde{%
\Psi}/}\left\vert \mathbf{\Psi }\right\vert ^{2}$ where $\mathbf{\tilde{\Psi}%
=}\left( \Psi _{x},\Psi _{y},\Psi _{z}\right) ^{T}$ \cite{Machida}. For
later use, we define the unit vector $\mathbf{s}\left( \mathbf{r}\right) =%
\mathbf{S}\left( \mathbf{r}\right) /\left\vert \mathbf{S}\left( \mathbf{r}%
\right) \right\vert $. 
\begin{figure}[htbp]
\begin{center}
\includegraphics[width=8.7cm]{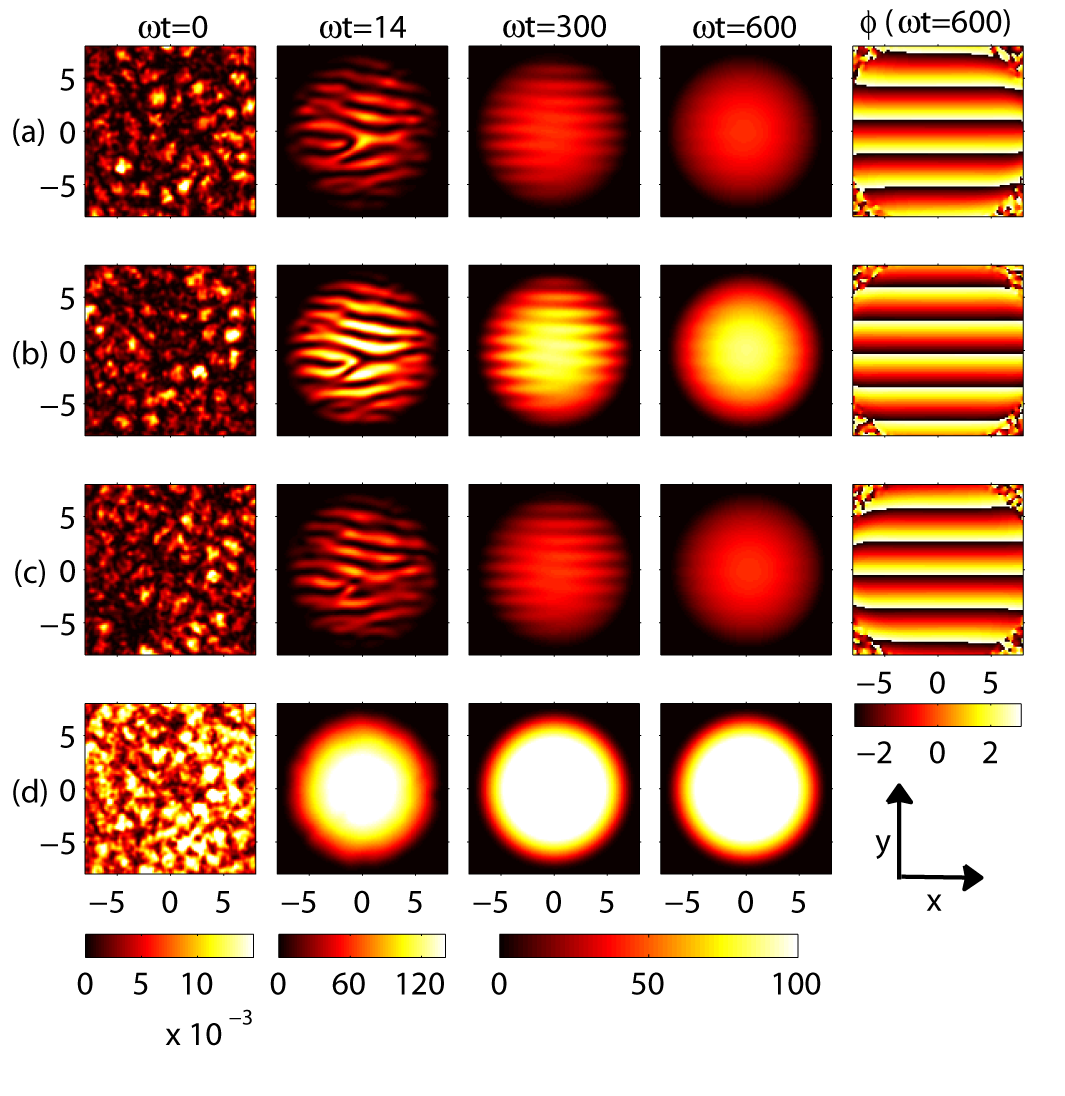}
\end{center}
\caption{(color online) Snapshots of (a) $\left\vert \Psi _{-1}\right\vert
^{2},$ (b) $\left\vert \Psi _{0}\right\vert ^{2},$ (c) $\left\vert \Psi
_{1}\right\vert ^{2},$ and (d) $\left\vert \mathbf{\Psi }\right\vert ^{2}$
of an $^{87}$Rb (FM) spinor BEC during quench with $V_{x}=1$, $V_{y}=2.$ The
rightmost column shows the phase profile of equilibrium state. The particle
numbers in the equilibrium state are $N_{\pm 1}\approx 3.65\times 10^{3},$ $%
N_{0}=7.32\times 10^{3}$.}
\label{Fig.1}
\end{figure}
\begin{figure}[htbp]
\begin{center}
\includegraphics[width=8.7cm]{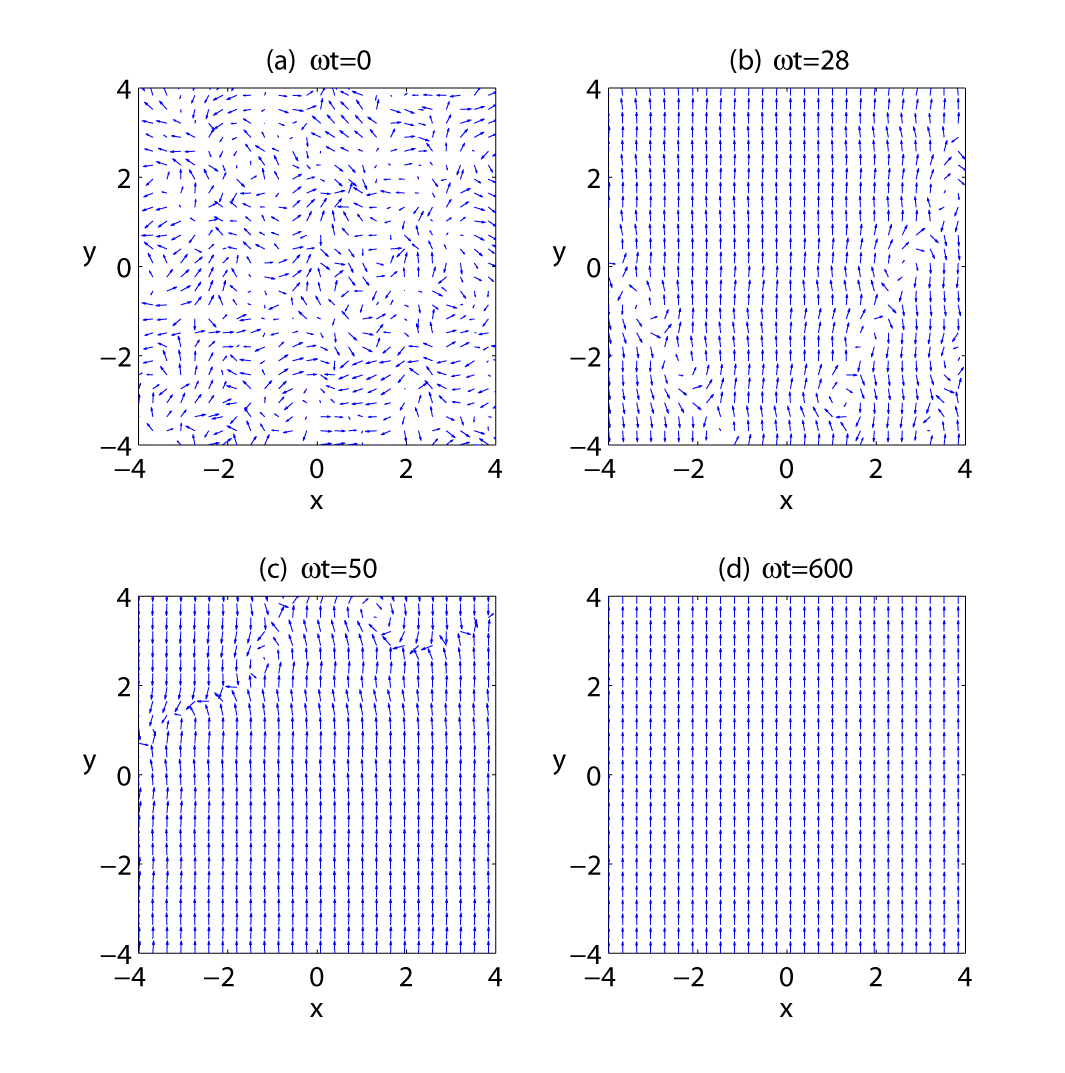}
\end{center}
\caption{Snapshots of the vector field $\mathbf{s}\left( \mathbf{r}\right) $%
. Arrays of merons of the same handedness can be seen in (b). All merons
drift away towards the periphery of the condensate as shown in (c) and (d).}
\label{Fig.2}
\end{figure}
In the mean-field theory approximation, the dynamics of a BEC at nonzero
temperatures can be described by the SPGPE\textit{\ }\cite{A. S. Bradley,
Blair, Blair2}, based on the assumption that the system can be treated as a
condensate band in contact with a thermal reservoir comprising of all
non-condensed particles. The condensate band is described by the truncated
Wigner method \cite{Sinatra} including the projected $c$-field method, while
the non-condensate band is by the quantum kinetic theory \cite%
{Gardiner1,Gardiner2}. A direct generalization to the spinor BECs with SOC
leads to the following set of coupled SPGPEs \cite{Su} 
\begin{equation}
\frac{\partial \Psi _{j}}{\partial t}=\mathcal{P}\{-\frac{i}{\hbar }\hat{H}%
_{j}^{GP}\Psi _{j}\mathbf{+}\frac{\gamma _{j}}{k_{B}T}\left( \mu -\hat{H}%
_{j}^{GP}\right) \Psi _{j}\mathbf{+}\frac{dW_{j}}{dt}\}  \label{SGPE}
\end{equation}%
where $T$ and $\mu $ denote the final temperature and chemical potential, $%
\gamma _{j}$ the growth rate for the $j$-th component, and $dW_{j}/dt$ is
the complex-valued white noise\ associated with the condensate growth. The
projection operator $\mathcal{P}$ restricts the dynamics of the spinor BEC
in the lower energy region below the cutoff energy $E_{R}$. For simplicity,
we shall consider a 2-dimensional isotropic trap, $U(\mathbf{r})=m\omega
^{2}\left( x^{2}+y^{2}\right) /2$. The numerical procedures for solving
SPGPEs, are described as follows. First, the initial state of each $\Psi
_{j} $ is sampled by using the grand-canonical ensemble for free ideal Bose
gas at a temperature $T_{0}$ below the critical temperature and of chemical
potentials $\mu _{j,0}$. The spatial dependence of the initial state can be
specified as a linear combination of plane waves with discretized momentum $%
\mathbf{k}=2\pi \left( n_{x},n_{y}\right) /L$ ($n_{x}$, $n_{y}$ $\in Z$ and $%
L$ is the size of the system), i.e., $\Psi _{j}\left( t=0\right) =\sum_{%
\mathbf{k}}^{E_{R}}a_{j,\mathbf{k}}\psi _{\mathbf{k}}\left( \mathbf{r}%
\right) $, where $\psi _{\mathbf{k}}\left( \mathbf{r}\right) =e^{i\mathbf{k}%
\cdot \mathbf{r}}$. The condensate band lies below the energy cutoff $%
E_{R}>E_{\mathbf{k}}=\hbar ^{2}\left\vert \mathbf{k}\right\vert ^{2}/2m$.
The distribution is sampled by $a_{j,\mathbf{k}}=\left( N_{j,\mathbf{k}%
}+1/2\right) ^{1/2}\eta _{j,\mathbf{k}}$ where $N_{j,\mathbf{k}}=\left[ \exp
(\left( E_{j,\mathbf{k}}-\mu _{j.0}\right) /k_{B}T_{0})-1\right] ^{-1}$ and $%
\eta _{j,\mathbf{k}}$ are the complex Gaussian random variables with moments 
$\left\langle \eta _{j,\mathbf{k}}\eta _{j,\mathbf{k}^{\prime
}}\right\rangle =\left\langle \eta _{j,\mathbf{k}}^{\ast }\eta _{j,\mathbf{k}%
^{\prime }}^{\ast }\right\rangle =0$ and $\left\langle \eta _{j,\mathbf{k}%
}\eta _{j,\mathbf{k}^{\prime }}^{\ast }\right\rangle =\delta _{\mathbf{kk}%
^{\prime }}$. Second, to simulate the thermal quench, the temperature and
chemical potential of the non-condensate band are altered to the new values $%
T<T_{0}$ and $\mu >\mu _{j,0}$. For convenience, we adopt the oscillator
units in the numerical computations, and the length, time and energy are
respectively scaled in units of $\sqrt{\hbar /m\omega }$, $\omega ^{-1}$ and 
$\hbar \omega $. 
\begin{figure}[htbp]
\begin{center}
\includegraphics[width=8.7cm]{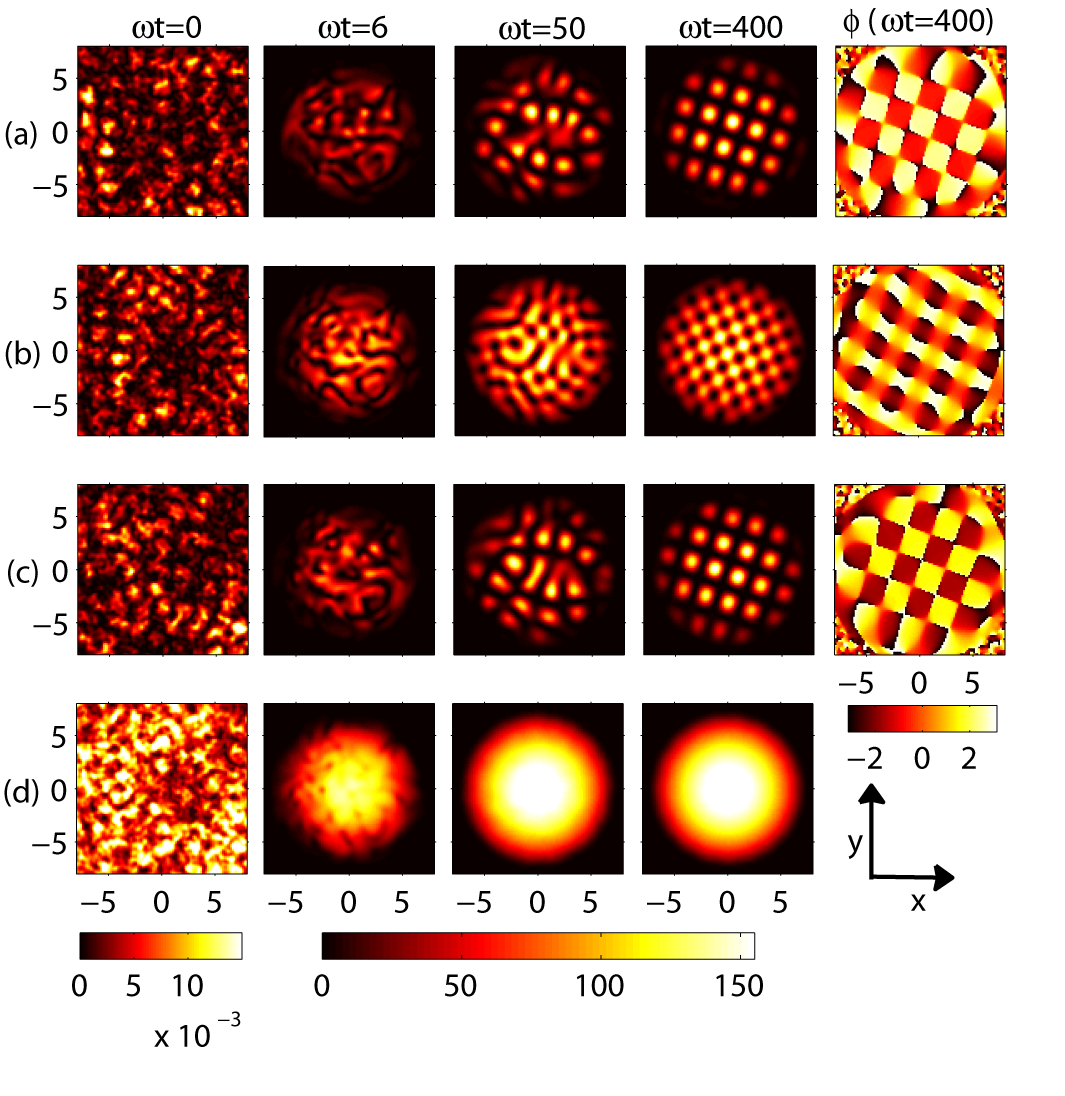}
\end{center}
\caption{(color online) Snapshots of (a) $\left\vert \Psi _{-1}\right\vert
^{2},$ (b) $\left\vert \Psi _{0}\right\vert ^{2},$ (c) $\left\vert \Psi
_{1}\right\vert ^{2},$ and (d) $\left\vert \mathbf{\Psi }\right\vert ^{2}$
of the $^{87}$Rb (FM) spinor BEC at zero magnetization with $V=1.8.$ The
rightmost column shows the phase profile of equilibrium state. The
corresponding particle numbers are $N_{\pm 1}\approx 3.56\times 10^{3},$ $%
N_{0}=7.08\times 10^{3}$.}
\label{Fig.3}
\end{figure}
\begin{figure}[htbp]
\begin{center}
\includegraphics[width=8.7cm]{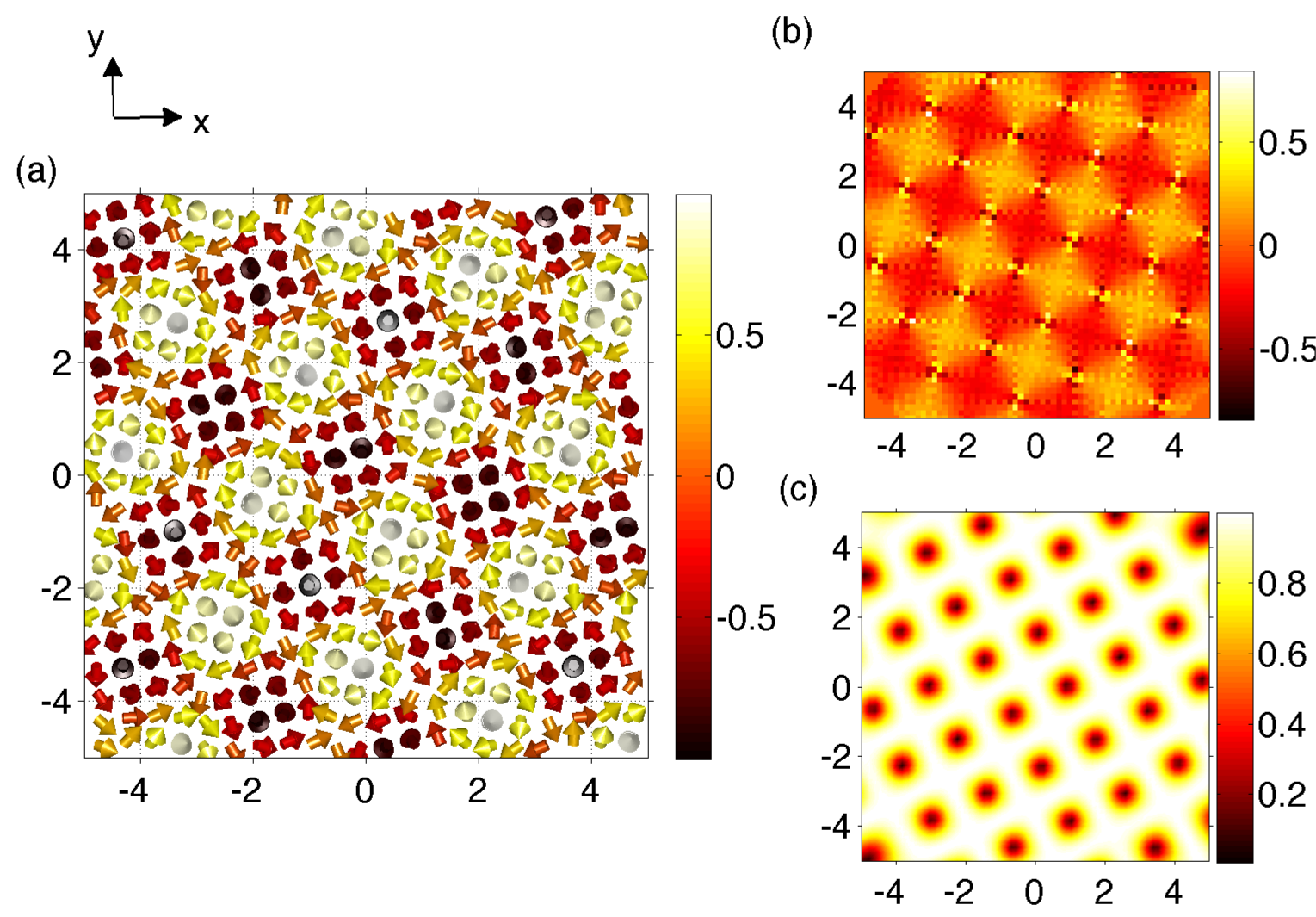}
\end{center}
\caption{(Color online) (a) The 3-dimensional orientation of the spin
textures in Fig. 1. The color indicate the magnitude of $s_{z}$. (b) The
topological charge density $\protect\sigma $ of the equilibrium spin
texture. (c) The corresponding spatial distribution of $\left\vert \mathbf{S}%
(\mathbf{r})\right\vert .$}
\label{Fig.4}
\end{figure}
We first study the condensation of $^{87}$Rb ($g_{s}<0$) with SOC. The
initial states are sampled at $k_{B}T_{0}=500$, $\mu _{j,0}=2$. Here we set $%
E_{R}=50.5$, $k_{B}T=0.067$, $\mu =25$, and$\ \hbar \gamma _{j}/k_{B}T=0.05$%
. In this paper, we consider the in-plane coupling, i.e., $V_{x},V_{y}\neq 0$
and $V_{z}=0$. We begin with the cases of $\left\vert V_{x}\right\vert \neq
\left\vert V_{y}\right\vert $. In Fig. 1, stripe structure develops in the
density profile of each component during condensation, but fade away
gradually. On the other hand, the equilibrium phase profiles become
periodically striped along one direction, which is is exactly the plane wave
(PW) state found in Ref. \cite{Wang}. The evolution of the spin texture is
shown in Fig. 2. Clearly, spin domains form shortly after the quench starts.
The domain walls comprise of sheets of spin vortex which drift outward and
disappear while the system approaches equilibrium marked by a uniform spin
alignment on the $xy$ plane. The PW state is bound to occur when $\left\vert
V_{x}\right\vert \neq \left\vert V_{y}\right\vert $, or $\left\vert
V_{x}\right\vert =\left\vert V_{y}\right\vert =V<0.8$. When $\left\vert
V_{x}\right\vert =\left\vert V_{y}\right\vert =V>0.8$, periodic structures,
which are caused by the formation of grids of dark soliton in $\Psi _{\pm 1}$
and vortex lattice in $\Psi _{0}$, may appear in the equilibrium density
profiles of all components in Fig. 3. In Fig. 4(a)-4(b), the spin texture
consists of two interlacing square lattices of spin vortex with opposite
vorticity, and we denote it as the spin vortex lattice (SVL) state. When $%
V>0.8$, PW and SVL state appear alternatively with definite probabilities
depending on the magnitude of $V$. Using the imaginary time propagation
method, the SVL state is shown to have a higher energy than the PW state. In
the SVL state, all spin vortices have similar structure --- the central spin
always orients to the $z$ axis, while the others increasingly twist and
finally lie on the $xy$ plane, forming a circulation pattern away from the
center. The topological charge density, $\sigma =\mathbf{s}\cdot \left(
\partial _{x}\mathbf{s}\times \partial _{y}\mathbf{s}\right) /4\pi $, is
plotted in Fig. 4(b) and the vortex with left/right-handed circulation has a
positive/negative $\sigma $. This is just the Mermin-Ho vortex, or \emph{%
meron} \cite{Mermin}. Integrating $Q=\int \sigma d^{2}r$ over a primitive
unit cell, we identify $Q=\pm 1/2$, corresponding to merons and antimerons,
respectively.\textbf{\ }Due to the FM nature of the condensate, a meron and
an antimeron will pair up to form a vortex dipole, which was predicted in
the bilayer quantum Hall system as the lowest energy excitation \cite{Brown}
and in a fast-rotating highly spin-polarized spinor BEC \cite{Machida} and
now acts as the building block of the SVL. For conciseness, in the
following, we shall not distinguish the merons from antimerons unless
specially noted.

To gain more insight into the SVL state, we plot the equilibrium momentum
wave functions, $\tilde{\Psi}_{i}\left( \mathbf{p}\right) $, and find that
they are all sharply peaked at $\mathbf{p}\approx \pm 2.8\mathbf{e}_{x}$, $%
\pm 2.8\mathbf{e}_{y}$. This suggests that the equilibrium state might be
consisted of 4 plane waves with $\mathbf{p=\pm q},\mathbf{\pm q}^{\prime }$,
where $\mathbf{q}\cdot \mathbf{q}^{\prime }=0$ and $\left\vert \mathbf{q}%
\right\vert =\left\vert \mathbf{q}^{\prime }\right\vert $. The
counter-propagating modes with $\mathbf{p=\pm q}$, $\mathbf{\pm q}^{\prime }$
form two orthogonal standing waves that superimpose to generate the periodic
structures in all density profiles in Fig. 3. In the absence of trapping
potential, it was shown that, the one-particle Hamiltonian, $p^{2}/2m+V%
\mathbf{p}\cdot \mathbf{F}$,\ is minimized by $\left\vert \mathbf{p}%
\right\vert =V$ \cite{Wang,You}, and hence 
\begin{eqnarray}
&&\tilde{\Psi}_{\pm 1}\left( \mathbf{p}\right) =\frac{\sqrt{N}e^{\mp i\theta
}}{4}\left[ -\delta \left( \mathbf{p-}\frac{V\mathbf{q}}{\left\vert \mathbf{q%
}\right\vert }\right) -\delta \left( \mathbf{p+}\frac{V\mathbf{q}}{%
\left\vert \mathbf{q}\right\vert }\right) \right.   \label{psi-1-t} \\
&&\left. \mp \delta \left( \mathbf{p-}\frac{V\mathbf{q}^{\prime }}{%
\left\vert \mathbf{q}^{\prime }\right\vert }\right) \mp \delta \left( 
\mathbf{p+}\frac{V\mathbf{q}^{\prime }}{\left\vert \mathbf{q}^{\prime
}\right\vert }\right) \right] ,  \nonumber \\
&&\tilde{\Psi}_{0}\left( \mathbf{p}\right) =\sqrt{\frac{N}{8}}\left[ \delta
\left( \mathbf{p-}\frac{V\mathbf{q}}{\left\vert \mathbf{q}\right\vert }%
\right) -\delta \left( \mathbf{p+}\frac{V\mathbf{q}}{\left\vert \mathbf{q}%
\right\vert }\right) \right.   \label{psi-0-t} \\
&&\left. +i\delta \left( \mathbf{p-}\frac{V\mathbf{q}^{\prime }}{\left\vert 
\mathbf{q}^{\prime }\right\vert }\right) -i\delta \left( \mathbf{p+}\frac{V%
\mathbf{q}^{\prime }}{\left\vert \mathbf{q}^{\prime }\right\vert }\right) %
\right] ,  \nonumber
\end{eqnarray}%
where $\theta =\tan ^{-1}\left( q_{y}/q_{x}\right) $.\ Consequently, the
spin textures are given by, 
\begin{eqnarray}
&&\mathbf{S}\left( \mathbf{r}\right) =\left( \cos \theta \cos u\sin v+\sin
\theta \sin u\cos v,\right.   \nonumber  \label{xx} \\
&&\left. \cos \theta \sin u\cos v-\sin \theta \cos u\sin v,\cos u\cos
v\right)   \label{SVL-spin}
\end{eqnarray}%
where $u=V(x\cos \theta +y\sin \theta )$ and $v=V(-x\sin \theta +y\cos
\theta ).$\ The above analytical results fairly reproduce the spin texture
plotted in Fig. 4(a) with an effective $V$\ in the presence of trapping
potential. From Fig. 4(c), we see that the spin density vanishes at the
center of a vortex quadruple (4 mutually adjoining merons), which happens to
be the density troughs of $\Psi _{\pm 1}$ that is filled by the particles of 
$\Psi _{0}$. The meron and polar cores are centered at $\mathbf{r}_{\text{%
meron/polar}}\mathbf{=}R\left( \theta \right) \mathbf{d}_{\text{meron/polar}}
$ where $R\left( \theta \right) $ is the rotation matrix on $xy$ plane, $%
\mathbf{d}_{\text{meron}}=\left( n,l\right) \pi /V$, $\mathbf{d}_{\text{polar%
}}=\left( n+1/2,l+1/2\right) \pi /V$,$\ $and $n,l\in Z$. It should be noted
that the cores of the meron are located at the vortices of $\Psi _{0}$. With
Eqs. (\ref{psi-1-t})-(\ref{psi-0-t}), the energies of PW and SVL states for
a homogeneous spinor BEC can be calculated analytically. As PW and SVL
states both yield the same minimized energy for the single particle
Hamiltonian with a fixed $N$, we only need to consider the interacting
energies. It is straightforward to show that both PW and SVL states lead to
the identical density-density interaction energies and thus the only
difference is the spin-exchange term. As a result, the spin-exchange
interaction energies per unit area, $\epsilon _{spin}=g_{s}N\int S^{2}\left( 
\mathbf{r}\right) d^{2}r$, are $g_{s}N$\ and $3g_{s}N/4$\ for PW and SVL
states, respectively. In the current case, $g_{s}<0$, and thus the SVL state
has a higher energy.

The SVL state is sustained by the vorticity originated mostly from SOC. As $%
\mu $ rather than $N$ is fixed in our simulations, increasing $V$ will
increase $N$ and the mass vortices in $\Psi _{0}$. Since PW and SVL states
are gapped by an amount of energy proportion to $N$, it is expected that
there exists a threshold $V_{c}$, beyond which the SVL will barely appear.
This implies that the PW state totally prevails in the large $V$ limit. The
value of $V_{c}$ can be estimated by considering a homogeneous spinor BEC,
in which the core size of the mass vortex nucleated in the $m_{F}=0$
component is $\xi _{0}\sim \hbar /\left\vert \Psi _{0}\right\vert \sqrt{%
2mg_{n}}$. The instability of SVL state occurs when the lattice constant is
comparable to $\xi _{0}$, such that a polar core will partially overlap with
the cores of contiguous merons, giving $V_{c}\sim \pi \left\vert \Psi
_{0}\right\vert \sqrt{2mg_{n}}/\hbar $. By the same token, if the size of
the condensate is smaller than the lattice constant, the condensate does not
accommodate SVL, which occurs when $V<0.8$ in our previous result in a trap.
In the limit of vanishing SOC, the spatial periodicity in the phase profiles
of PW and SVL states is infinitely prolonged, leading to uniform
distribution in all phase profiles --- the manifestation of a normal spinor
BEC. 
\begin{figure}[htbp]
\begin{center}
\includegraphics[width=8.7cm]{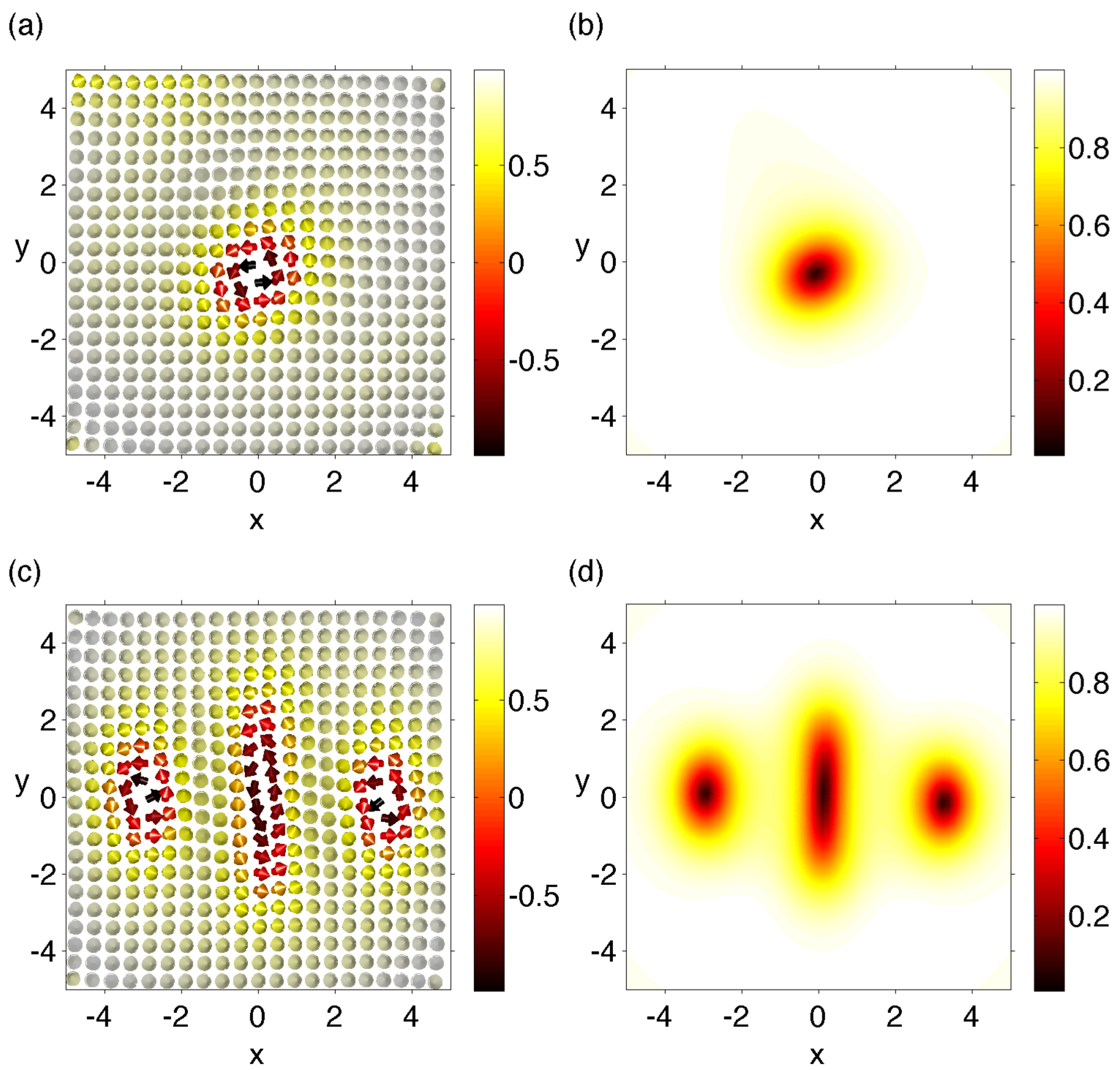}
\end{center}
\caption{(color online) Spin textures and the associated distribution of $%
\left\vert S\left( \mathbf{r}\right) \right\vert $ in a $^{23}$Na (AFM)
spinor BEC for\ the inverted merons with (a), (b) $M/N=0.88$; (c), (d) $%
M/N=0.78$. The particle numbers in the equilibrium state are $N_{1}\approx
4.66\times 10^{4},$ $N_{0}=5.14\times 10^{3},$ $N_{-1}\approx 3.74\times
10^{2}$ in (a) and $N_{1}\approx 4.18\times 10^{4},$ $N_{0}=9.51\times
10^{3},$ $N_{-1}\approx 9.03\times 10^{2}$.}
\label{Fig.5}
\end{figure}
\begin{figure}[htbp]
\begin{center}
\includegraphics[width=8.7cm]{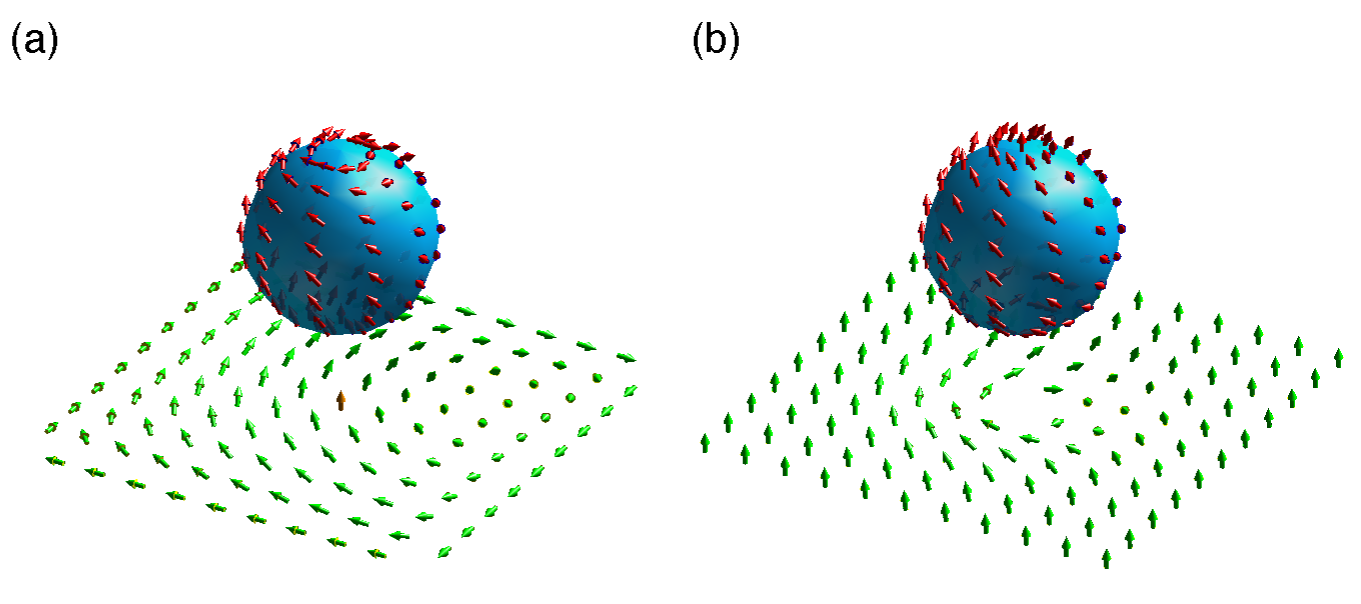}
\end{center}
\caption{(Color online) (a) The stereographic projection of the spin
textures onto $S^{2}$ for (a) a meron; (b) an inverted meron.}
\label{Fig.6}
\end{figure}
We now consider the spinor BEC of \ $^{23}$Na ($g_{s}>0$) with SOC. Using
the same parameter setting, we conclude that if the spin-1 gas is
unpolarized initially, no spin texture will be produced in the condensate, a
conclusion already given in Ref. \cite{Wang}. The equilibrium spin texture
for a spin-polarized state with $M/N\approx 0.88$ is shown in Fig. 5(a),
where all spins in the outer region point to the $+z$ direction, tilting
gradually, and finally lie on the $xy$ plane while approaching the center
region. Consequently, a polar core is formed at the center, with $\Psi _{0}$
filling in the vortices of both $\Psi _{\pm 1}$ which are separated but
locate proximately to the center. Such a configuration corresponds to a
charge of $Q=-1/2$. In Fig. 6 the stereographic projections for a meron and
the texture of Fig. 5(a) are plotted. Clearly, Fig. 6(b) is exactly the
inverted figure of Fig. 6(a), and thus we term the structure of Fig. 5(a)
the \emph{inverted meron}. In Fig. 5(c)-5(d), three inverted merons with
distorted cores are formed with $M/N\approx 0.78$. It is easy to verify that
the single isolated inverted meron has a lower energy.

In summary, we have investigated the non-equilibrium dynamics of spin-1 BECs
with SOC in the limit of rapid quench. Crystallization of merons and polar
core vortices are predicted to arise in the FM spinor BEC. Likewise,
isolated inverted merons can be created in the highly polarized AFM spinor
BEC. Our studies provide a method to create nontrivial structure of merons
and thus an opportunity to probe into the fundamental properties of
meron-like matter. Following the experimental methods in Ref.\cite{Lin}, our
predictions can be realized in principle, except in lifting the degeneracy
of the hyperfine spin states of $F=1$, a weaker magnetic field is needed to
avoid the decoupling of spin states due to quadratic Zeeman shift. The
experiment may start by trapping a thermal spin-1 Bose gas and then rapidly
quench it to the quantum degenerate regime. Finally, the equilibrium spin
texture can be resolved \emph{in situ} by using the polarization-dependent
phase-contrast technique \cite{Higbie}.

S.-C. Gou is supported by NSC under Grant No. 100-2112-M-018-001-MY3. W. M.
Liu is supported by NSFC under Grant No. 10934010, and NKBRSFC under Grant
No. 2011CB921502. S.-C. Gou thanks Dr. Y.-J. Lin and Dr. M.-S. Chang for the
helpful discussions.


\newpage

\begin{thebibliography}{99}
\bibitem{Sakurai} J. J. Sakurai, \textit{Modern Quantum Physics }%
(Addison-Wesley Publishing Company, Inc., U.S.A. 1994).

\bibitem{Lin} Y.-J. Lin, K. Jim\'{e}nez-Garcia and I. B. Spielman, Nature 
\textbf{471}, 83 (2011).

\bibitem{Liu} X.-J. Liu \textit{et al}., Phys. Rev. Lett. \textbf{102},
046402 (2009).

\bibitem{Bandy} S. Bandyopadhyay, Phys. Rev. B \textbf{61}, 13813 (2000).

\bibitem{Das} I. \v{Z}uti\'{c}, J. Fabian, and S. Das Sarma, Rev. Mod. Phys. 
\textbf{76}, 323 (2004).

\bibitem{Sinova} J. Sinova \textit{et al}., Phys. Rev. Lett. \textbf{92},
126603 (2004).

\bibitem{Bernevig} B. A. Bernevig, T. L. Hughes, and S.-C. Zhang, Science 
\textbf{314}, 1757 (2006).

\bibitem{Zhang} Y. Zhang \textit{et al}., Nature Phys. \textbf{6}, 584
(2010).

\bibitem{Kane} M. Z. Hasan and C. L. Kane, Rev. Mod. Phys. \textbf{82}, 3045
(2010).

\bibitem{Hui} H. Hu, H. Pu, and X.-J. Liu, eprint arXiv:1108.4233v1.

\bibitem{Wang} C. Wang \textit{et al}., Phys. Rev. Lett. \textbf{105},
160403 (2010).

\bibitem{You} Z. F. Xu, R. L\"{u}, and L. You, Phys. Rev. A \textbf{83},
053602 (2011).

\bibitem{Kawakami} T. Kawakami, T. Mizushima, and K. Machida, Phys. Rev. A 
\textbf{84}, 011607(R) (2011).

\bibitem{Yi} W. Yi and G.-C. Guo, Phys. Rev. A \textbf{84}, 031608(R) (2011).

\bibitem{Jiang} L. Jiang \textit{et al}., eprint arXiv:1110.0805v1.

\bibitem{Jayantha} J. P. Vyasanakere and V. B. Shenoy, eprint
arXiv:1108.4872v1.

\bibitem{Kibble} T. W. B. Kibble, J. Phys. A \textbf{9}, 1387 (1976).

\bibitem{Zurek} W. H. Zurek, Nature \textbf{317}, 505 (1985).

\bibitem{Actor} A. Actor, Rev. Mod. Phys. \textbf{51}, 461 (1979).

\bibitem{Brey} L. Brey \textit{et al}., Phys. Rev. B \textbf{54}, 16888
(1996).

\bibitem{Brown} G. E Brown and M. Rho, \textit{The Multifaced Skyrmion}
(World Scientific Singapore, 2010).

\bibitem{Ruutu} V. M. H. Ruutu, \textit{et al}., Phys. Rev. Lett. \textbf{79}%
, 5058 (1997).

\bibitem{Ishiguro} R. Ishiguro, \textit{et al}., Phys. Rev. Lett. \textbf{93}%
, 125301 (2004).

\bibitem{Ketterle} A. E. Leanhardt \textit{et al}., Phys. Rev. Lett. \textbf{%
90}, 140403 (2003).

\bibitem{Ho} T.-L. Ho, Phys. Rev. Lett. \textbf{81}, 742(1998).\textbf{\ }

\bibitem{Ohmi} T. Ohmi and K. Machida, J. Phys. Soc. Jpn. \textbf{67}, 1822
(1998).

\bibitem{Machida} T. Mizushima \textit{et al}., Phys. Rev. A \textbf{70},
043613 (2004).

\bibitem{A. S. Bradley} A. S. Bradley \textit{et al}., Phys. Rev. A \textbf{%
77}, 033616 (2008).

\bibitem{Blair} S. J. Rooney, A. S. Bradley, and P. B. Blakie, Phys. Rev. A 
\textbf{81}, 023630 (2010).

\bibitem{Blair2} S. J. Rooney \textit{et al}., Phys. Rev. A \textbf{84},
023637 (2011).

\bibitem{Sinatra} A. Sinatra \textit{et al}., J. Phys. B \textbf{35}, 3599
(2002).

\bibitem{Gardiner1} C. W. Gardiner \textit{et al}., Phys. Rev. A \textbf{58}%
, 1050 (1998).

\bibitem{Gardiner2} C. W. Gardiner \textit{et al}., Phys. Rev. A \textbf{61}%
, 033601 (2000).

\bibitem{Su} S.-W. Su, \textit{et al}., Phys. Rev. A \textbf{84}, 023601
(2011).

\bibitem{Mermin} N. D. Mermin, \textit{et al}., Phys. Rev. Lett. \textbf{36}%
, 594 (1976).

\bibitem{Higbie} J. M. Higbie \textit{et al}., Phys. Rev. Lett. \textbf{95},
050401 (2005).\newpage
\end{thebibliography}
\end{document}